# Scheimpflug cameras for range-resolved observations of the atmospheric effects on laser propagation

Nathan Meraz[1,*], Megan Birch[1], Ian Winski[1,2], Mary Kate Broadway[1,3],
Denly Lindeman[1,2], Katie Twitchell[1,2], and Joseph L. Greene[1];

[1] Georgia Tech Research Institute, Electro-Optical Systems Lab, Atlanta, GA, 30332, USA
[2] University of Arizona, Wyant College of Optical Sciences, Tucson, AZ, 85721, USA
[3] Georgia Institute of Technology, College of Engineering, Atlanta, GA, 30332, USA
*Nathan.Meraz@gtri.gatech.edu

## ABSTRACT

This paper presents the development of Scheimpflug cameras for lidar and remote sensing with an emphasis on active and passive range-finding. Scheimpflug technology uses a tilted camera geometry to natively encode 3D information through projected off-axis pixel view angles and holds the unique potential to serve as an alternative to traditional lidar and remote sensing systems with the demonstrated advantages of high configurability, SWaP-C (Size, Weight and Power-Cost) efficiency, and short- vs. far-range optimization. In this work, we demonstrate a compact Scheimpflug-enabled system as a snapshot atmospheric lidar detector to measure aerosol extinction and optical turbulence effects with high precision over ranges from a few meters to a few kilometers. We compare the instrument's measurements to variance-based Cn2 data collected by a sonic anemometer and a scintillometer over a 50 m horizontal path. This paper also presents preliminary results on utilizing Scheimpflug technology for photogrammetry, 2D/3D mapping and includes a generalized discussion on design, alignment, and calibration procedures. We believe this work provides a strong basis for the broad use of Scheimpflug technology across multiple use-cases within the fields of lidar and remote sensing.

## 1. INTRODUCTION

Lidar systems are widely employed in remote sensing, atmospheric monitoring, and industrial diagnostics due to their spatial and temporal resolution and sensitivity to environmental changes along a direct line of sight. Traditional lidar configurations rely on time-of-flight methods, which require fast detectors, high-speed timing and digitizing electronics, and precisely pulsed or modulated lasers to resolve range information. As engineered instruments, these systems are challenging due to their inherently interconnected link budgets and constrained design spaces. Improving one aspect of the system often requires modifications across multiple subsystems to maintain efficiency and operational effectiveness. Consequently, lidar engineers have reduced ability to balance performance metrics such as total range, resolution, sensitivity within functional constraints like SWaP-C and system complexity. The desire for a flexible approach to both lidar design and performance optimization is our motivation for exploring alternatives. Here, we leverage the Scheimpflug principle [1], to provide a more adaptable and efficient ranging approach that allows for greater variability of the performance parameters and fundamentally more options for sensor hardware. We believe that exploring alternatives like this will motivate new holistic approaches to lidar and remote sensing tasks while creating new design spaces with beneficial tradeoffs to near-range applications.

The Scheimpflug principle provides a means of optical triangulation where the planes of the image, optics, and object are tilted to intersect along a common line. This configuration allows an imaging system to extend the in-focus object plane over a wider span of depths and geometrically map range information directly to the sensor coordinates. Scheimpflug triangulation has historically found use in optical metrology due to the enhanced depth resolution and extended imaging volume [2][3]. More recently, the methods have coincided modern with structured light techniques for photogrammetry and 3D profiling [4][5][6] and utilize lasers as illuminators to isolate in-focus target features and aid the reconstruction processing. The application to lidar was demonstrated clearly by Mei and Brydegaard [7][8], who introduced camera-based Scheimpflug lidar systems for atmospheric aerosol profiling and differential absorption measurements featuring continuous wave (CW) lasers and CCD sensors. Many more versions of these systems have since been developed as reduced SWaP-C alternatives to time-of-flight lidar. The applications have included scattering classification measurements within a few meters [9][10][11][12], variations of atmospheric profiling over horizontal and vertical paths [13][14][15][16][17][18][19], and tracking and classification [20]. Many of these efforts derive the environmental measurement based on the scattering intensity by modeling the propagation of the illumination laser and estimating the sampling volume for each camera pixel. The substantial body of available research validates the fundamental approach of Scheimpflug lidar (S-lidars) and it is feasible and promising to utilize these systems to enhance laser propagation evaluations and assess near-range optical turbulence.

Precise evaluation of laser propagation is vital in various domains, including remote sensing, laser communication, and directed energy systems where the atmospheric extinction, scattering, and turbulence can significantly attenuate and distort the optical field. The spatial and temporal variation of refractive index, referred to as optical turbulence, leads to the specific issues of beam wander, beam spread, scintillation, and wavefront distortion that are often associated with optical system performance. Evaluating, predicting, and mitigating performance issues is aided by having a series of continuous measurements of the integrated effect of the atmospheric conditions along the path. Time-of-flight lidar is a typical technique for single-ended atmospheric profiling and is particularly useful for long-range measurements. However, time-of-flight lidars have practical limitations relating to dynamic range, timing complexity, calibration, and range resolution. As a result, they can be inadequate for making profiled measurements over very short ranges, such as near the output aperture of a laser transmitter where turbulence has more influence on the down range beam. Turbulence lidars that do not instantaneously measure the extended volume of the beam path [21] are likely under sampled spatially and temporally.

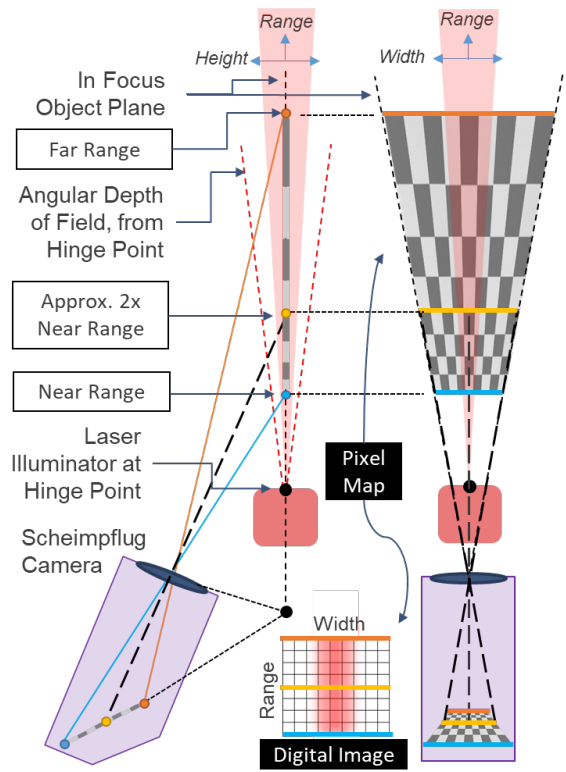

Figure 1: Diagram of the Scheimpflug lidar concept with laser aligned to object plane. Pixel columns are mapped to range and pixel rows are mapped to width in the plane. Left side is a view of the Range-Height plane showing the relative tilts of the camera lens, sensor, and object plane and the projection of column pixels to Range. These project along chief rays through the center of the camera lens. Right side shows view normal to the object plane where the width of the beam is sampled by row pixels at each range. The representative output image shows no variation in beam width when the pixel field of view and beam divergence are constant with range.

In this paper, we show that images captured by Scheimpflug lidars, configured as in Figure 1, can be used for direct evaluation of laser propagation by estimating beam width and beam wander. We discuss the development and testing of Scheimpflug lidars that purposefully circumvent the issues of time-of-flight lidar with comparable performance as a short-range instrument (< 2 km) while also offering SWaP-C advantages. This work also includes experiments to observe turbulence effects on beam propagation characteristics such as beam width and scintillation, with measurements supported by a sonic anemometer and scintillometer. An additional experiment uses a UV-sensitive Scheimpflug camera to capture vertical atmospheric data simultaneously with a typical elastic time-of-flight lidar.

## 2. RESEARCH FRAMEWORK

The primary objectives of our research are centered around the design, development, and validation of a Scheimpflug camera system for lidar applications. First, we derive parameterized imaging equations and construct a camera integrated with a 532 nm laser illuminator to determine the optical ranging capabilities. This derivation also includes detailing a calibration process to ensure accurate distance mapping. Then, with the laser aiming at a scintillometer, we conduct controlled experiments to observe the system's sensitivity to optical turbulence. The scintillometer indicates the integrated effect of optical turbulence along the beam path in the forward direction, which is referenced to the range-profiled back/side scatter measurements made by the Scheimpflug lidar. We repeat the setup outdoors along with a sonic anemometer to provide point measurements of Cn2 to verify the scintillation index values calculated from the scintillometer.

Additionally, we develop a UV-sensitive Scheimpflug camera and operate it simultaneously with a traditional elastic lidar. We compare the atmospheric extinction data collected by both systems and attempt to correct the range-dependent volume sampling of the Scheimpflug lidar.

### 2.1 Scheimpflug Principle

The conventional thin lens imaging equations rely on central projection and collinear transformations to map points, lines, and planes between object and image space relative to a lens plane and its optical axis. The Scheimpflug principle states that object points on a plane tilted amount $\theta$ will project to points on a plane tilted by amount $\theta'$ if those planes intersect at the principal planes of the lens, as illustrated in Figure 2.

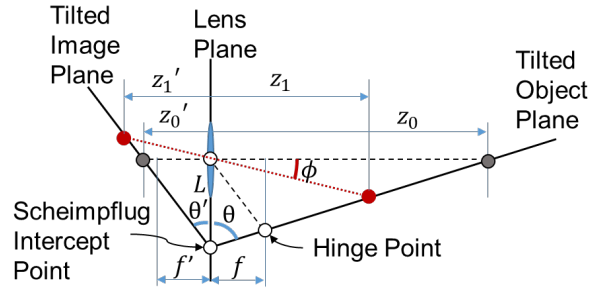

Figure 2: Scheimpflug imaging geometry using conventional Gaussian imaging notation. Axial distances z and z' are defined from the lens plane which contain the principal points. The lens with front and rear focal lengths $f$ and $f'$ is shown imaging two different points between the tilted object and image planes. These planes intersect with the lens plane a distance L from the optical axis. The hinge point is at the intersection of the object plane and front focal plane. [22]

In relation to the customary Gaussian imaging equation,

$$\frac{1}{z'} + \frac{1}{z} = \frac{1}{f} \qquad (1)$$

the expressions for optical magnification and line of sight angle succinctly state the Scheimpflug principle:

$$m_n \equiv \left(\frac{z_n'}{z_n}\right)\left(\frac{f}{f'}\right) = \frac{\left(\frac{\sin\theta'}{\cos(\theta'+\phi_n)}\right)}{\left(\frac{\sin\theta}{\cos(\theta-\phi_n)}\right)} \tag{2}$$

$$\tan\phi_n = \frac{z_0 - z_n}{z_n \tan\theta} \tag{3}$$

Additionally, the tilt angles $\theta$ and $\theta'$ are defined by the right triangles formed by the on-axis distances $z_0$ and $z_0'$ and the Scheimpflug baseline distance $L$. However, the full expression relating points in the object and image planes is relatively complex in this notation [7].

### 2.2 Scintillation Index and Cn2 Values

The range-resolved scintillation index, which quantifies the fluctuations in the intensity of the laser beam due to atmospheric turbulence, is calculated to estimate the Cn2 values along the beam path. Fluctuation of irradiance $I$ are commonly characterized by the scintillation index expression

$$\sigma_I^2 = \frac{\langle I^2 \rangle - \langle I \rangle^2}{\langle I \rangle^2} \tag{4}$$

which is the normalized variance of the fluctuations. When the scintillation index is less than 1, the fluctuations are considered "weak" and the variance in the beam is considered directly proportional to the Rytov variance of an ideal plane wave:

$$\sigma_I^2 = 1.23 C_n^2 k^{\frac{7}{6}} L^{\frac{11}{6}} \tag{5}$$

The Cn2 value is the refractive-index structure constant and is a measure of the strength of optical turbulence in the atmosphere. Together, these metrics are used to evaluate laser propagation and relate optical turbulence to observable changes to intensity.[25] As illustrated in Figure 1, the S-lidar is capable of capturing sequential images of the propagating beam and provides sampled data for making range profiles of beam statistics. The S-lidar's range resolution, angular resolution, exposure time, bit-depth, and frame rate, laser power and wavelength are design variables that can be optimized to facilitate specific measurement requirements.

## 3. RESULTS

### 3.1 Scheimpflug Lidar Ranging Equations

Developing a Scheimpflug camera for lidar requires a method for mapping image coordinates to ranges and pixel numbers to their associated range bins. To meet this need, we propose a robust geometric ranging model based on Newton's imaging equation that provides a straightforward relationship for the ranging distance $R$ and the sensor array coordinate $R'$.

Newton's equation and the updated definition of magnification:

$$NN' = ff' \tag{6}$$

$$m \equiv \frac{N'}{f'} = \frac{f}{N} \tag{7}$$

use distances relative to the focal planes of the optical system located at $f$ and $f'$ from the lens plane, as shown in Figure 3. If the index of refraction is identical in object and image space then $f' = f$.

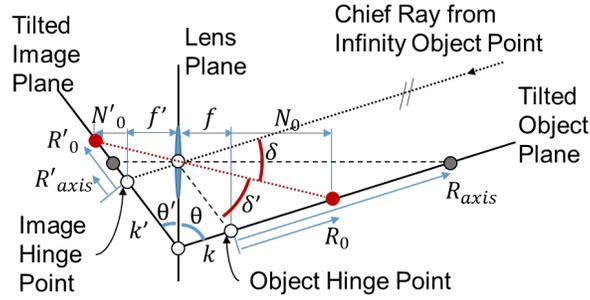

Figure 3: Scheimpflug imaging geometry using Newton imaging notation. Axial object and image distances N and N' are defined from the front and rear focal planes. Distances R and R' along the tilted planes are measured from the hinge points. These planes intersect with the lens plane a distance L from the optical axis.

The distances $N$ and $N'$ represent the object and image distances, but are measured along the optical axis. The simplicity of Newton's equation can be maintained by projecting the optical axis onto the object and image planes. The front and rear focal lengths project to lengths $k$ and $k'$:

$$k = \left(\frac{f}{\sin(\theta)}\right) \tag{8}$$

$$k' = \left(\frac{f'}{\sin(\theta')}\right) \tag{9}$$

and the conjugate axial object and image distances project to $R$ and $R'$:

$$R = \left(\frac{N}{\sin(\theta)}\right) \tag{10}$$

$$R' = \left(\frac{N'}{\sin(\theta')}\right) \tag{11}$$

This projection creates new expressions for Newton's equations but does not alter the underlying definitions from (6) and (7). The Scheimpflug range equation and associated magnification ratio:

$$RR' = kk' \tag{12}$$

$$m = \frac{R'}{k'} = \frac{k}{R} \tag{13}$$

provide these simple expressions for $R$ and $R'$:

$$R = \frac{k}{m} = \frac{kk'}{R'} \tag{14}$$

$$R' = mk' = \frac{kk'}{R} \tag{15}$$

This set of equations describes the mapping of points between the sensor plane and the object plane with a simple relationship between the controllable design values: object plane angle $\theta$, image plane angle $\theta'$, and the focal length $f$.

Next, we seek to define range sampling resolution $\Delta R$ in relation to the discrete pixel size $\Delta R'$. Since we are using 2D sensor arrays with consistent pixel size, $\Delta R'$ is a well-defined design constant. Evaluating the derivative of (14) or (15) gives the Scheimpflug ranging equation's instantaneous rate of change:

$$\frac{\delta R'}{\delta R} = \frac{-kk'}{(R)^2} = \frac{-(R')^2}{kk'} \tag{16}$$

The incremental change using discrete step sizes can be approximated to use a single range:

$$\frac{\Delta R'}{\Delta R} = \frac{-kk'}{R_2 R_1} \approx \frac{-kk'}{(R_1)^2} \tag{17}$$

or in a more useful form to evaluate range resolution $\Delta R$ using sensor coordinates:

$$\frac{\Delta R}{\Delta R'} = \frac{-kk'}{R'_2 R'_1} \approx \frac{-kk'}{(R'_1)^2} \tag{18}$$

These can also be re-written using just range and sensor distances which is useful for calibrations:

$$\Delta R \approx -\Delta R' \frac{R}{R'} \tag{19}$$

These equations use the camera's optical parameters to determine the geometrical minimum and maximum range and the range resolution for each pixel. The defined imaging relationships are required for calibration and used to evaluate depth of field and effects to the general lidar equation (22). In particular, (17) shows that each pixel (with fixed image length $\Delta R'$) has a projected range depth of $\Delta R$ that increases with the square of the range $R$.

The approximation in (17)-(19) assumes that the change in magnification associated with $\Delta R$ or $\Delta R'$ is small enough to consider being constant. In general, this is valid when $\frac{\Delta R}{R} \ll 1$ and $\frac{\Delta R'}{R'} \ll 1$ but will often not be the case when trying to extend the maximum object range to infinity causing the magnification $m$ and $R'$ approach zero.

## 3.2 Alignment and Range Calibration of a Scheimpflug Lidar

Calibration is a fundamental step in ensuring the accuracy and reliability of the range measurements. The process involves adjusting the hardware or model values to provide a known relationship between the observed values and the true or desired values.

The goal for calibration is to make two or more calculations of magnification. A key property of the Scheimpflug geometry is that the magnification changes linearly with pixel index as indicated by (15) and demonstrated in Figure 4. Additionally, the magnification is 0 at the pixel containing the infinity object image. If the Scheimpflug geometry is not perfectly configured or there are uncertainties in the calibration or target placement, a best fit line can be found with linear regression following the magnification expression defined by (13).

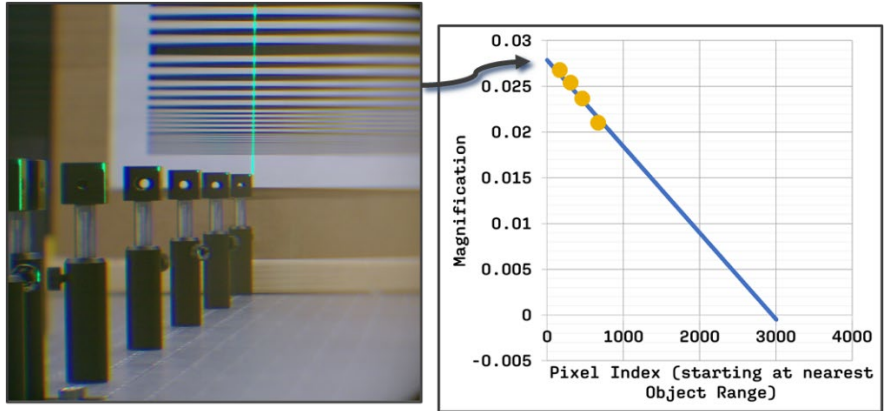

Figure 4: (Left) Example calibration image with multiple targets with known separations. (Right) Corresponding map of magnification vs. pixel index.

The first step involves aligning the camera sensor to ensure that it accurately maps the pixels in the sensor plane to an object plane extending from the camera. The adjustments shown in Figure 5 are necessary to align all the axes and points to image a fixed laser across a specific row of pixels. The alignment requires having targets in the desired object plane at known ranges. After initial alignment to get both targets in focus and at least one of the targets in the correct column a calibration image of the targets is taken. The pixel indices of the target images are used to get uncalibrated $R$ values using the initial design parameters for comparison to the known $R$ values. The calculation of the measured magnification for each conjugate object and image pair determines the as-built Scheimpflug parameters and magnification map. This process is aided by equation (19). Instead of using a target at known distances from the object hinge, it is possible to instead use three or more targets with known separations between them.

A calibrated system with known values for $k$, $k'$, and the $R'$ sensor coordinates can use equation (14) as a sensor-to-range map. Additionally, (14) and (15) show that ranging errors stem from magnification errors. Thus, the optics used in a real system should ideally minimize aberrations affecting magnification to preserve the straightforward geometrical relationships described previously. Otherwise calibration measurements can provide an "as-built" map of magnification vs. pixel index that strays from the expected linear relationship.

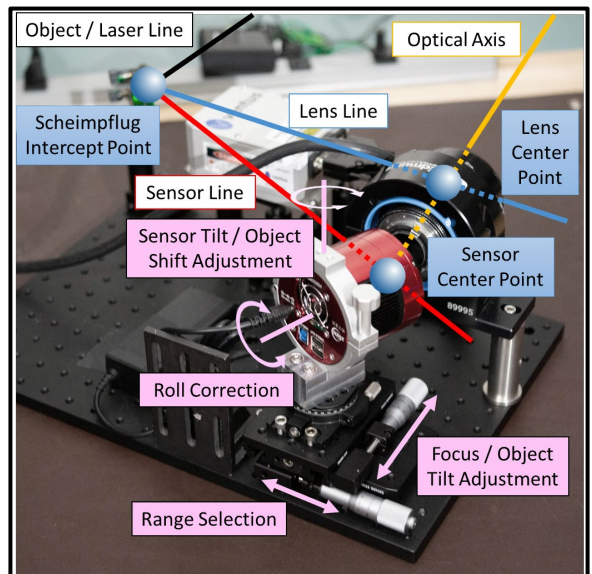

Figure 5: Diagram showing the optomechanical adjustments needed for alignment. In later prototypes, these adjustments were removed in favor of 3D printed and machined components with minimal adjustments that position the sensor and lens close to the nominal design positions. For those systems, the calibration procedures were used to measure the "as-built" ranging parameters.

If the calibrated values (resolution and ranges) don't match the desired model but the object plane is in good focus, it indicates that the offset of the camera from the object plane (the Scheimpflug baseline distance $L$) is incorrect. Achieving the desired minimum and maximum range requires translating the entire camera perpendicular to the lens' optical axis to re-position the Scheimpflug point. Moving along that line keeps the hinge point within the desired object plane. Making this adjustment changes the angle of the best focus plane which requires either re-adjustment of the focus to rotate the object plane about the hinge point or an adjustment of the laser's propagation direction.

The laser illuminator needs to be aligned coincident with the object plane. Reflecting the beam off a mirror at the hinge point gives the ability to align the laser and keep it in the desired plane. Knowing the beam's propagation properties (initial size and divergence) is crucial to properly measure the effect on the beam size as it propagates in different environmental conditions.

### 3.3 Prototype Overview

A single camera can provide range information passively in accordance to the geometrical imaging relationships determined by the Scheimpflug principle. By controlling the camera sensor's position and tilt angle, it is possible to precisely map sensor coordinates to regions of an object plane tilted away from the camera. This concept has a variety of applications because of the versatility of camera systems in terms of sensors, wavelengths, optics, and image processing techniques. A major benefit is that the Scheimpflug technique works with any lens and sensor array and can be configured to work over different ranges and resolutions by adjusting the component positions and tilts.

The prototypes for this research were originally designed for researching laser propagation and optical turbulence, but our efforts have extended into surface roughness mapping and 3D point clouds to demonstrate the broad generality of Scheimpflug design. Figure 1 shows the conceptual arrangement of the Scheimpflug cameras we developed into lidars. Most applications require an illumination laser but it is possible to use the camera as a passive system if image processing algorithms are able to isolate "in-focus" features in the object plane. The system's ranging is straightforward to calibrate and maintains similar resolutions and radiometric performance as standard cameras using the same components. Simple geometric relationships define the preliminary design and range performance needed for system engineering. Two of these systems, shown in Figure 6, were used to analyze beam propagation and estimate optical turbulence. The prototypes used the components in Table 1 and have the ranging maps shown in Figure 7, with image examples in Figure 8 and Figure 9.

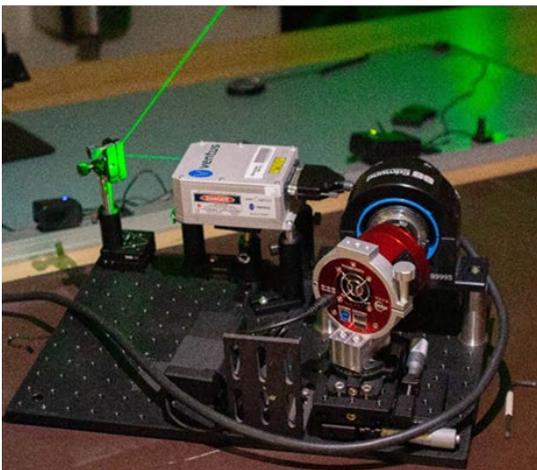 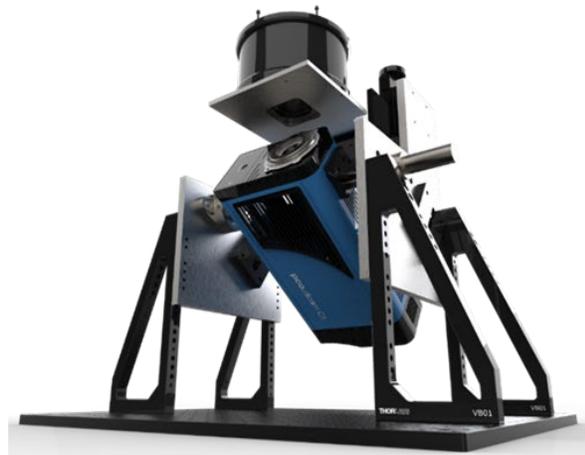

Figure 6: (Left) "532 nm Prototype" with 1.5W CW illumination laser used for controlled experimentation and rooftop horizontal measurements. Operational range 5 – 200 m. (Right) Rendering of "355 nm Prototype" built to simultaneously image an elastic lidar's 355 nm laser. It utilizes a range-gated intensified camera shown below a 6" F/3 telescope which was later swapped for a 200 mm lens for an operational range of 5 – 1500 m.

Table 1: Key Hardware specifications for the prototype Scheimpflug lidars

| Components | 532 nm Prototype | 355 nm prototype |
|---|---|---|
| Camera Lens | Nikon F/2.8, 135 mm focal length, 48 mm aperture diameter | 200 mm focal length, 48 mm aperture diameter. |
| Camera Sensor | ZWO ASI533MM Pro, Sony IMX533, 11.3 x 11.3 mm, 3.76 μm pixels, 3008 x 3008, monochrome, 16bit, TEC cooling | pco.dicam c1 intensified 16 bit sCMOS (DICAMC125s20-P46). Intensifier: 6 μm pixels, 25 mm diameter, 0.53:1 relay magnification. sCMOS: 2048x2048, 6.5 μm pixels, 13.3 x 13.3 mm (18.8 mm diagonal) |
| Laser | Ventus, 532 nm, 1.5 W, 1 mrad divergence | Continuum 9050, 355 nm, 200 mJ pulses at 50 Hz (10 W), 0.1 mrad divergence with beam expander |
| Narrowband Filter | Edmund Optics 65-216, 532 nm, 10 nm bandpass, OD4.0 | Edmund Optics 34-492, 355 nm, 10 nm bandpass, OD4.0 |
| Scheimpflug Geometry | Sensor tilt: 19.8°. Laser Plane tilt: 2.2°. Scheimpflug baseline: 0.38 m | Sensor tilt: 21.4°. Laser Plane tilt: 0.0°. Scheimpflug baseline: 0.51 m |

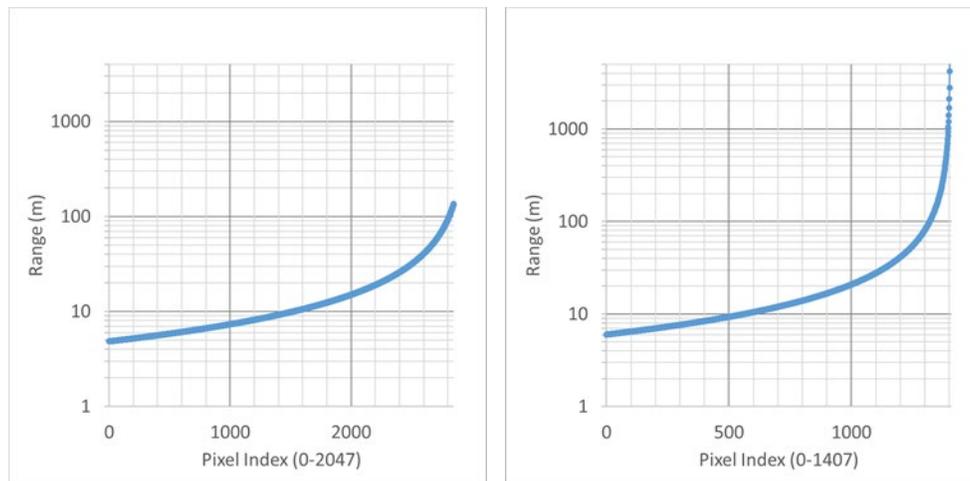

Figure 7: Pixel-to-range maps for idealized mapping of the sensor plane to a thin object plane. Left: 532 nm prototype. Right: 355 nm prototype

Like a regular camera, the radiometric power collected in each pixel when imaging large objects is approximately constant with range. When imaging the volumetric backscatter of a laser several pixels wide, these Scheimpflug lidars do not exhibit the characteristic range squared power issues of time-of-flight lidar. Effectively, the range squared change to the solid angle of the receiver from the source is negated by a range squared change to projected pixel area. While this is only exact for an infinitely thin object, it is more or less evident in the prototype systems. This means that when used as a lidar, the dynamic range and bit depth of the sensor is used to capture more detailed information. Additionally, background image collections are possible by having a modulated or pulsed laser. Such images can be subtracted from data frames to improve contrast and detectability and also be used to determine biased noise statistics in each pixel.

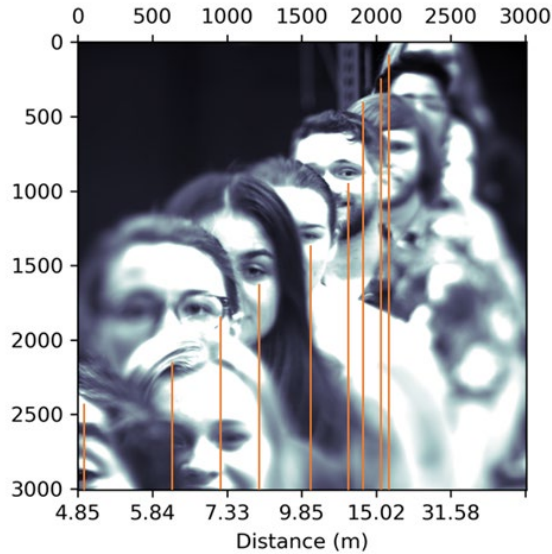

Figure 8: Image taken with 532 nm prototype to demonstrate mapping capabilities. Orange vertical lines indicate location of best focus for each individual in the image. Geometric pixel-to-range map applied to calculate distance scale.

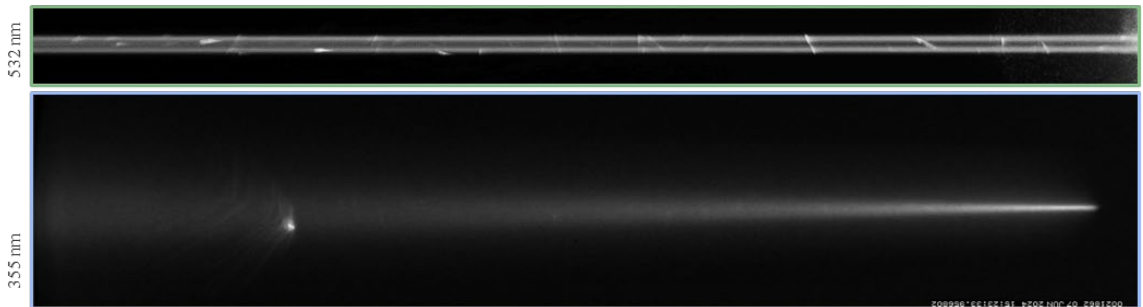

Figure 9: Laser beam images directly from the prototypes with beam propagating from left to right. (Top): 532 nm prototype with dual beam output. Range 5 - 15 m with 0.3 ms exposure. Horizontal propagation indoors with dust in the beam path. (Bottom): 355 nm prototype image of the elastic lidar's laser. Vertical propagation through the atmosphere with clear skies. Large aerosol or insect in beam path.

### 3.4 Theory of Passive and Active Scheimpflug Ranging

The nearest and farthest operational distances depend on the focal length and sensor height. The pixel pitch determines the range sampling resolution. Sliding the sensor along the image plane will adjust the nearest and farthest ranges. Adjusting the axial position effectively tilts the object plane. A longer focal length will narrow the field-of-view (FOV), reduce the operational ranges and increase the range resolution. Like other cameras, a Scheimpflug camera has a hyperfocal distance. In the context of ranging, point objects between the hyperfocal distance and infinity will image into the same range column. Figure 7 shows that when the camera is configured to image infinity, the middle pixel coordinate maps to approximately twice the nearest range. These maps are derived from the thin lens imaging equation and the triangulation geometry of the Scheimpflug principle.

### 3.5 Results in Passive Photogrammetry

Dimensional information of objects lying in or extending through the in-focus plane is calculated from the sampled range value and the angular image size. With good contrast along the object plane, a single image from the camera is enough to measure the length of features within that plane. The resolution of the measurement is tied to the angular FOV of the pixel and the distance to the object plane. Idealized performance plots for the 532 nm prototype system are shown in Figure 10.

At 100 m, the range resolution is 0.7 m with a 2.6 mm target size resolution. The FOV of the pixel's is calculated from the image distance and projected size of the pixel. Like other cameras, the geometric pixel FOV is subject to a tangent relationship but remains nearly constant for narrow FOV systems; it only changes about 6% across the approximately 4° FOV for this system.

Scheimpflug cameras have a depth of focus in image space limited by the working F/# and pixel size, but the tilted geometry results in an angular depth of field in object space. This means images of large objects with features off the in-focus plane have increasing amounts of defocus the further they extend from the plane. As a passive camera, determining the minimum amount of defocus is required to estimate the range bin at which a large object passes through the object plane.

Figure 11 shows a demonstration of the photogrammetry technique where a single image from the 532 nm prototype measured over 1000 points over a 2 m distance with sub-millimeter resolution.

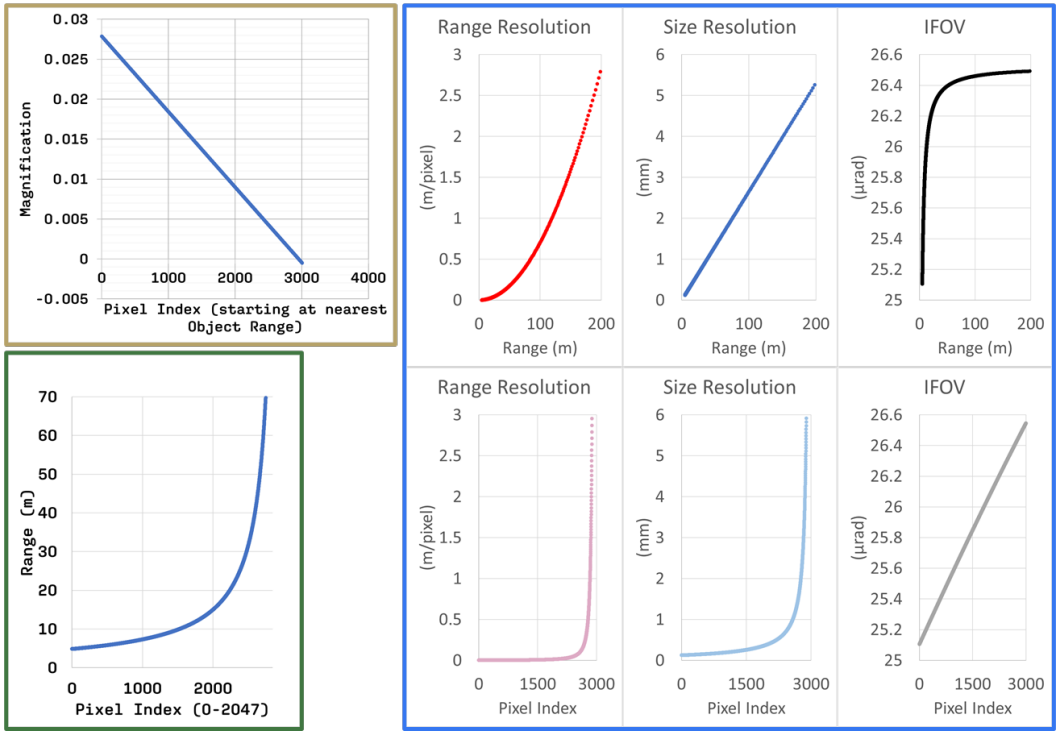

Figure 10: Design performance plots for the 532 nm prototype. (Top Left) Magnification-pixel map. (Bottom Left) Range-pixel map. (Right) Geometric resolutions versus range and versus pixel. These are idealized geometric values that do not include the depth of focus or diffraction effects.

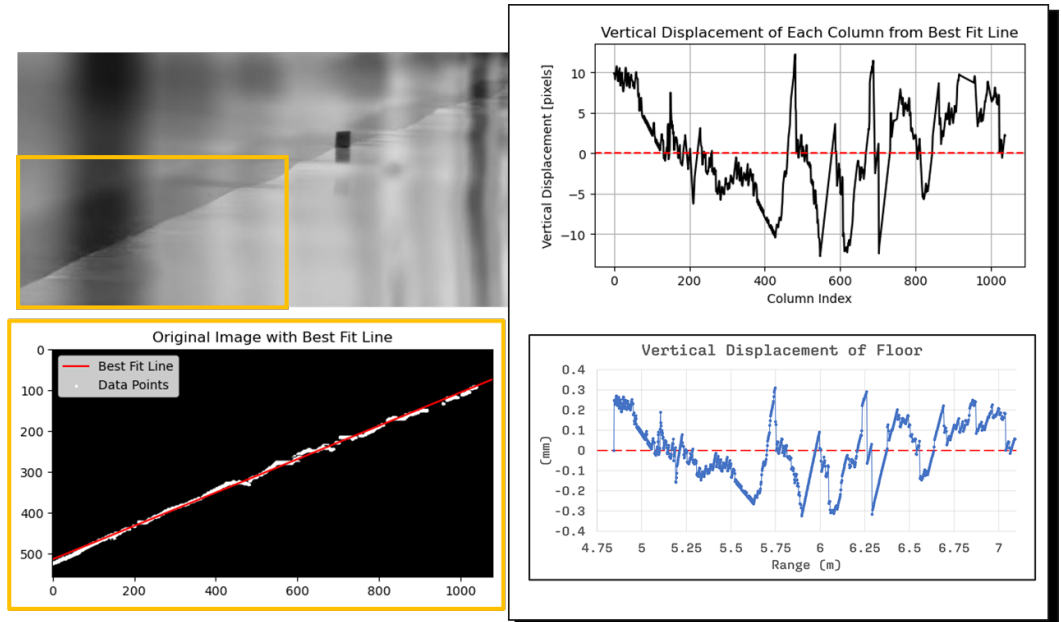

Figure 11: Photogrammetry example. Scheimpflug camera measurement of floor irregularity. (Top Left) The object plane of the camera was aligned to intersect the edge of a dark stripe on the floor. (Bottom Left) The collected image was processed to determine the in-focus edges of the stripe across 1000 range pixels. A best fit line was used as the reference for the average height. (Top Right) For each range column, the row offset was measured in units of pixels. (Bottom Right) The vertical displacement at each range was calculated using the system's range-pixel map and size resolution plot.

### 3.6 Utility of Laser Illumination for Active Scheimpflug Applications

Adding a laser illuminator (either pulsed or continuous) improves performance and provides more measurement capabilities. Background and foreground clutter relative to the object plane are naturally out of focus, but a laser illuminator aligned to the object plane simplifies range determination and improves ranging accuracy of hard targets. Measuring laser backscatter strength and beam image statistics (e.g., beam width and scintillation) over time can profile volumetric targets like aerosols and other atmospheric effects including optical turbulence. Estimating the size, speed, and direction of particulate matter crossing through the beam is also possible using images like in Figure 9.

### 3.7 Results in 2D and 3D Scanning

A Scheimpflug lidar array with $m$ rows and $n$ columns would have $n$ range samples along $m$ lines of sight. Illuminating each of line of sight for 2D coverage requires a laser "sheet", an array of lasers or angular scanning of a single laser to illuminate the extent of the object plane. For limited 3D coverage, angular scanning of the object plane is possible with linear axial translation of the sensor plane. For a given system with a fixed sensor tilt, all possible object planes will intersect at the object hinge point. A laser steering mirror placed in that location can keep the illumination laser within the object plane as it tilts. Rotating the sensor around the image hinge point will cause a linear translation of the object plane. These types of motions, shown in Figure 12, are also useful to for alignment.

Figure 13 shows the process for generating a 3D point cloud using a Scheimpflug lidar. The range-calibrated camera was placed on a 200 mm translation stage. A constant frame rate and stage motion was used to determine the relative position for each image frame. Range and height of features were calculated from the processed camera images. The range, height, and position values were used for object points in the 3D visualization.

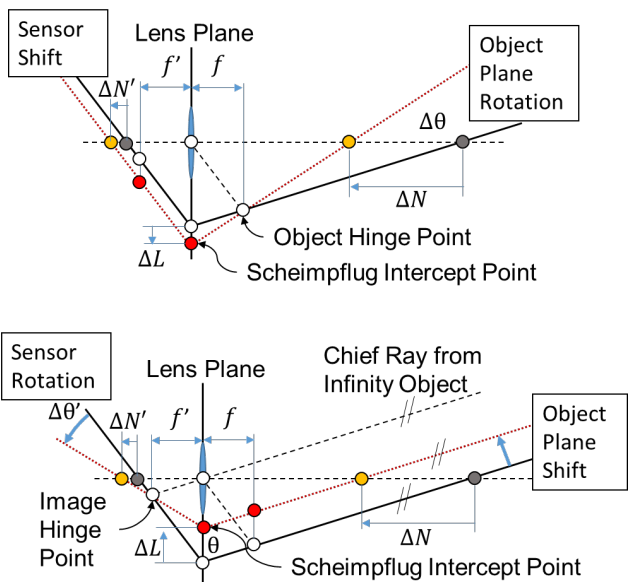

Figure 12: Scanning and alignment of image and object planes is achieved by repositioning the Scheimpflug intercept point along the lens plane. In both diagrams, the resultant axial object and image points are shifted by $\Delta N$ and $\Delta N'$. (Top) Axial shift of the sensor causes a shift of the Scheimpflug intercept point and a rotation of the object plane around the object hinge point (Bottom) Rotation of the sensor around the image hinge point causes a shift of the Scheimpflug intercept point and a translation of the object plane.

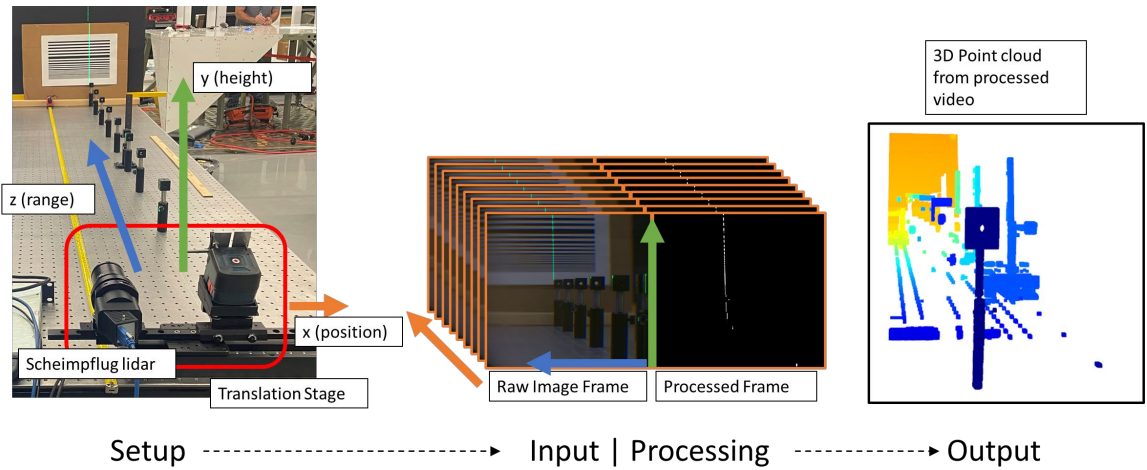

Figure 13: Preliminary process for generating a 3D point cloud. This setup used a small Scheimpflug camera and "sheet laser" to illuminate the entire object plane which natively generates range and height measurements. The lidar was attached to a translation stage to capture sequential frames from different scan positions. Each frame was processed to identify and label features as points with 3D coordinates to then form a 3D visualization with colors corresponding to depth from the camera's focal plane. The depth range spans nearly 6'.

### 3.8 Results in Optical Turbulence Experiments

The general concept for the turbulence measurements was to use the Scheimpflug lidar to measure the motion statistics in each range column, relate that motion to optical turbulence models, and calculate the Cn2 range profile. The experimental setup utilized a scintillometer to provide information for the optical turbulence model based on the measured scintillation index. Figure 14 shows an example of the process from initial validation efforts.

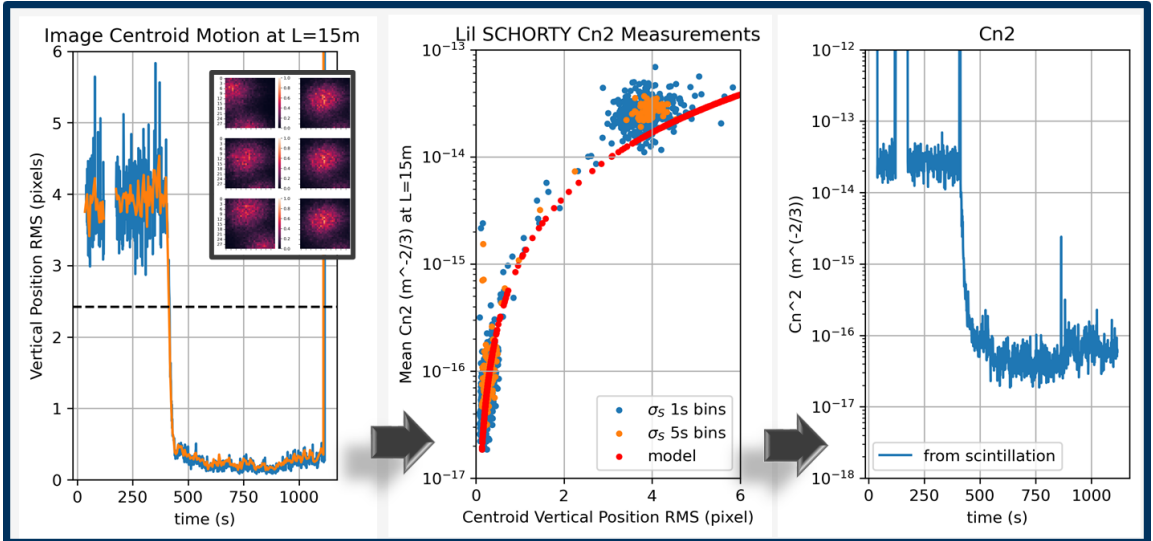

Figure 14: Optical turbulence measurement concept. Image motion statistics over time are used with a turbulence model to calculate Cn2. Example shows data from a controlled experiment with a hard target at 15 m rather than a full range profile. A fan blowing warm air into the beam path was disabled at 400 seconds, resulting in a sharp drop to image motion and Cn2.

The outdoor turbulence experiment was performed in Atlanta, Ga on 5/8/2023 from 6:45p to 9:00p to capture the sunset transition. An adjacent sonic anemometer measured wind and temperature for Cn2 calculations. The goal of the experiment was to compare the Scheimpflug lidar, scintillometer, and sonic anemometer. In particular, we expected to calibrate the scintillometer using the sonic anemometer and attempt range profiling with the Scheimpflug lidar. Figure 15 shows an unfiltered image taken by the Scheimpflug camera during setup. The beam was aimed into the scintillometer at a range of approximately 60 m. Figure 16 are examples of the filtered and cropped beam images from the dual beam configuration at different exposures.

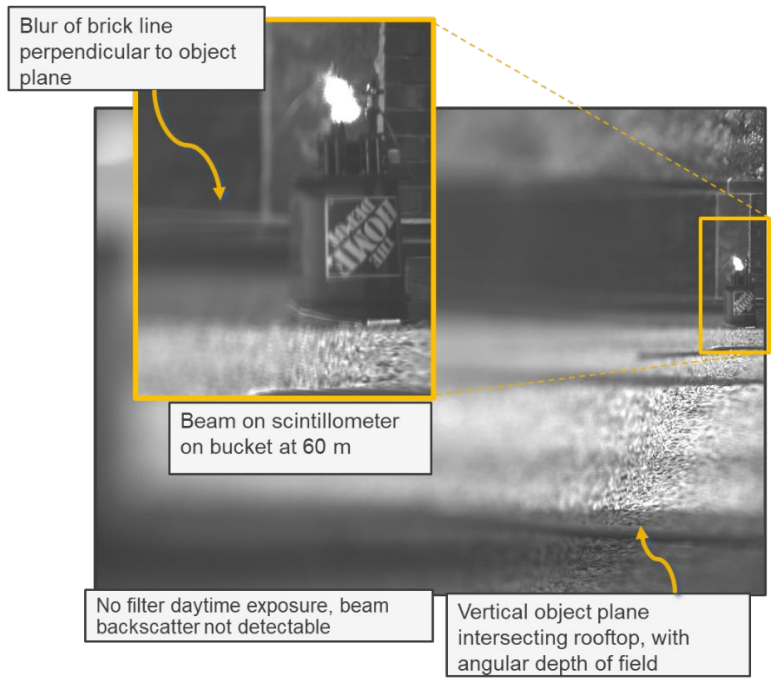

Figure 15: Unique features noticed in Scheimpflug images during the setup of the outdoor turbulence experiment with the 532 nm prototype

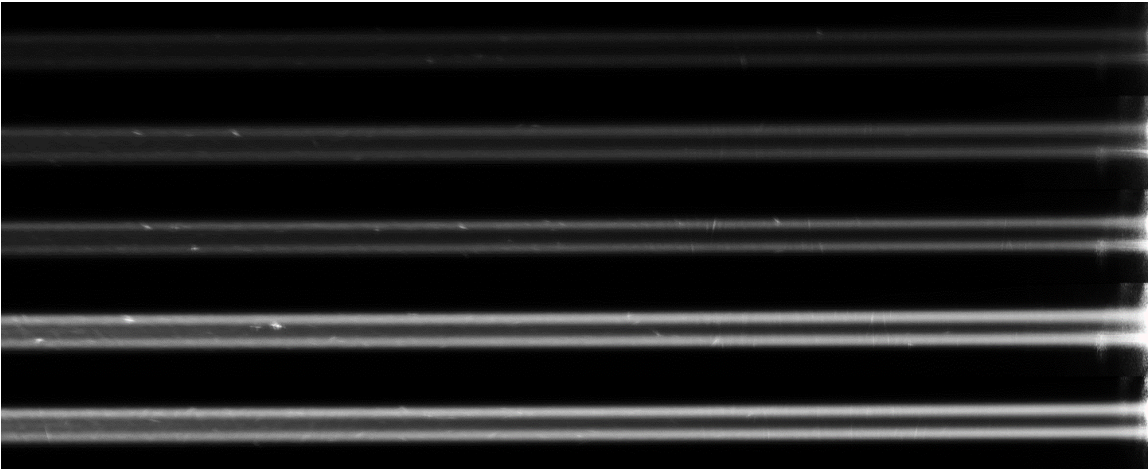

Figure 16: Beam image examples. Top to bottom exposure times are 1s, 2s, 3s, 5s, 8s. Images stretched vertically to show texture

We collected and processed the video data from the S-lidar and the scintillometer. Unfortunately, we discovered that we did not collect sufficient data to utilize the approach shown previously in Figure 14 so we did not fulfill our objective of making ranged profiles of Cn2. However, the S-lidar data was still adequate for calculating the beam width at each range. Figure 17We observed a downward trend to both Cn2 measurements along with the S-lidars beam width measurement.

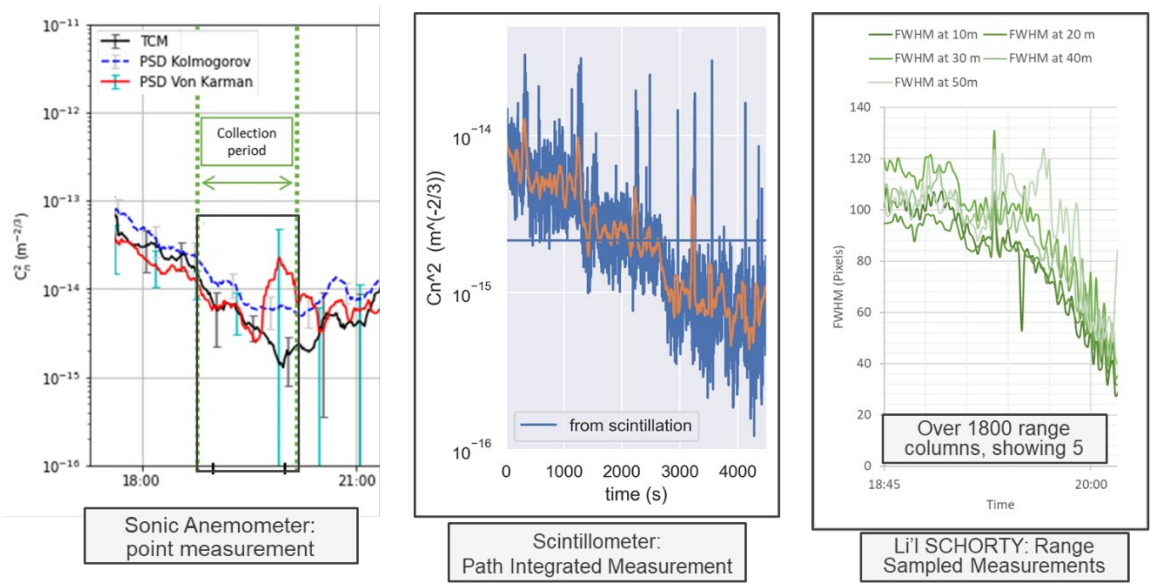

Figure 17: Calculated data from the outdoor turbulence experiment. Scintillometer value are uncorrected but still show a similar relative drop in Cn2 compared to the sonic anemometer over the collection period. A similar trend is seen in the FWHM beam width reported by the S-lidar. The plot is showing 5 of the over 1800 ranges between 10 m and 50 m.

## 3.9 Results and Theory for Atmospheric Extinction Characterization

Atmospheric extinction is another critical parameter for understanding light propagation and aerosol distributions in the atmosphere. On June 7, 2024 we used the UV 355 nm prototype Scheimpflug camera to make concurrent measurements with a custom time-of-flight elastic lidar. Example images are shown in Figure 18. The goal for this experiment was for preliminary observations and modeling, including an attempt at range correction. By imaging the lidar's laser beam, we were able to compare the range-corrected power and extinction measurements from both systems. The backscatter profiles in Figure 19 were created with minimal processing of the image data.

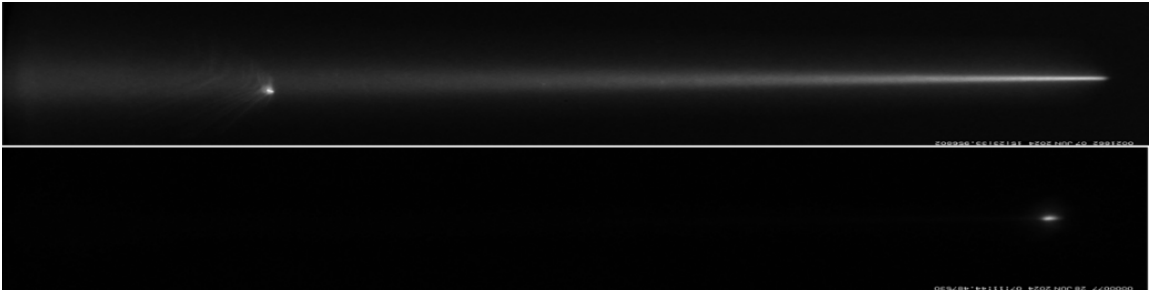

Figure 18: Images from the UV 355 nm Scheimpflug lidar. (Top) Beam image taken during normal operating conditions. A small insect in the beam path at around 10 m. (Bottom) Beam image taken at low power to observe scattering from low cloud (approximately 120 m).

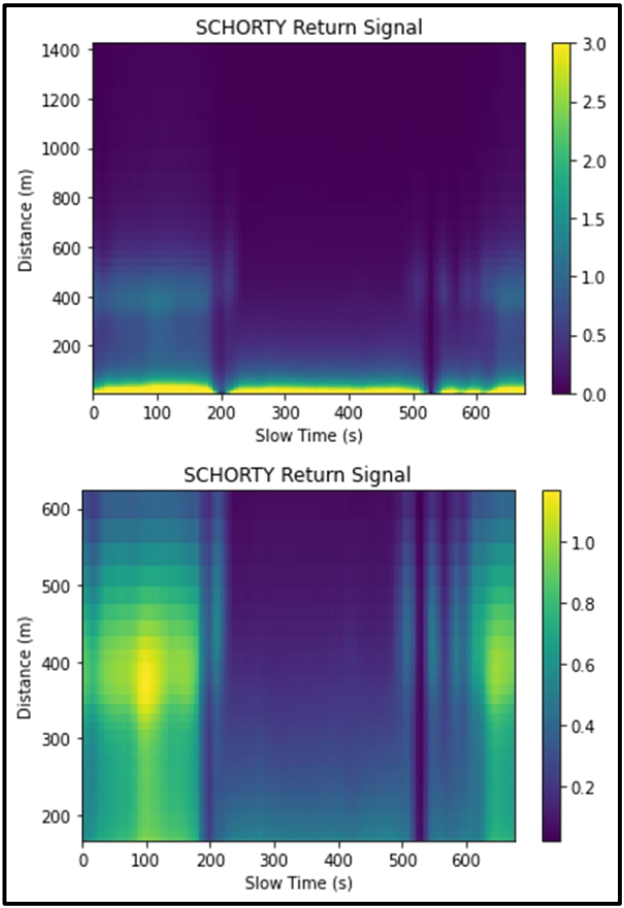

Figure 19: Backscatter profile from the UV 355 nm S-lidar over a 10-minute period. 50 Hz data reduced to 1 Hz. (Top) Full range plot covering 6m to 1400 m. (Bottom) Selection showing 180 m – 620 m with cloud layer at around 400 m. The images appear smeared in the range direction, likely due to the need for a volume sampling correction.

Using a Scheimpflug camera as a lidar requires an additional model to determine the sampling volume of each pixel. Additionally, due to the depth of field, the effective range resolution is larger than the geometric projection of the pixel onto the ideal object plane.

We developed a preliminary correction model using basic assumptions. We first assume that the beam width is narrower than the depth of field in order to ignore defocus effects. Thus, the extended range resolution is primarily related to the beam size as follows:

$$dz_{eff} = \frac{z(n_p) \cdot w(z(n_p))}{L} \tag{20}$$

where $L$ is the designed Scheimpflug baseline offset, $z$ is ranging distance, and $w$ is beam diameter. At far ranges past the Rayleigh range, the waist increases approximately linearly with distance such that:

$$dz_{eff} \approx \frac{\theta z^2(n_p)}{L} \tag{21}$$

where $\theta$ is the beam divergence.

The general lidar equation is:

$$P_s = \frac{K\beta(\lambda, z)}{z^2} dz \cdot \exp\left(-2 \int_0^z \alpha(\lambda, z') dz'\right) \tag{22}$$

For an ideal Scheimpflug systems considering only the object plane, $dz = z^2$ which would eliminate the range squared dependence. However, that isn't quite correct due to having a significantly finite sampling width, so we implement a correction by multiplying the lidar equation by a new term

$$\chi = \frac{dz_{eff}}{dz} \tag{23}$$

Past the Rayleigh range, this quantity reduces to $\chi = \frac{\theta}{L}$. Before the Rayleigh range, this term equals:

$$\chi = \frac{\frac{z(n_p) \cdot 2 \cdot w_0 \cdot \sqrt{1 + \left(\frac{\lambda z}{\pi w_0^2}\right)^2}}{L}}{z(n_p)^2} = \frac{2 \cdot w_0 \cdot \sqrt{1 + \left(\frac{\lambda z}{\pi w_0^2}\right)^2}}{L \cdot z(n_p)} \tag{24}$$

Our implementation of the ranging model used the approximation shown in equation (18) which makes an improper assumption regarding the sampling resolution of the pixels mapped to the farthest ranges. To try to account for the inaccuracy, we considered implementing an additional correction factor as part of the range correction:

$$\psi = \frac{z(n_p) \cdot z(n_p + 1)}{z(n_p)^2} \tag{25}$$

As a result, the total correction becomes:

$$\chi \cdot \psi = \frac{2 \cdot w_0 \cdot \sqrt{1 + \left(\frac{\lambda z}{\pi w_0^2}\right)^2}}{L \cdot z(n_p)} \cdot \frac{z(n_p) \cdot z(n_p + 1)}{z(n_p)^2} = \frac{z(n_p + 1) \cdot 2 \cdot w_0 \cdot \sqrt{1 + \left(\frac{\lambda z}{\pi w_0^2}\right)^2}}{z(n_p)^2} \tag{26}$$

Multiplying this by our data gives a range corrected signal. To then find the extinction term $\alpha$ in the lidar equation, we apply the slope method. This is done by taking the natural log, fitting a linear trendline to the linear region, and finding the slope. Dividing the slope by -2 gives the one-way extinction. The results of this process are shown in Figure 20.

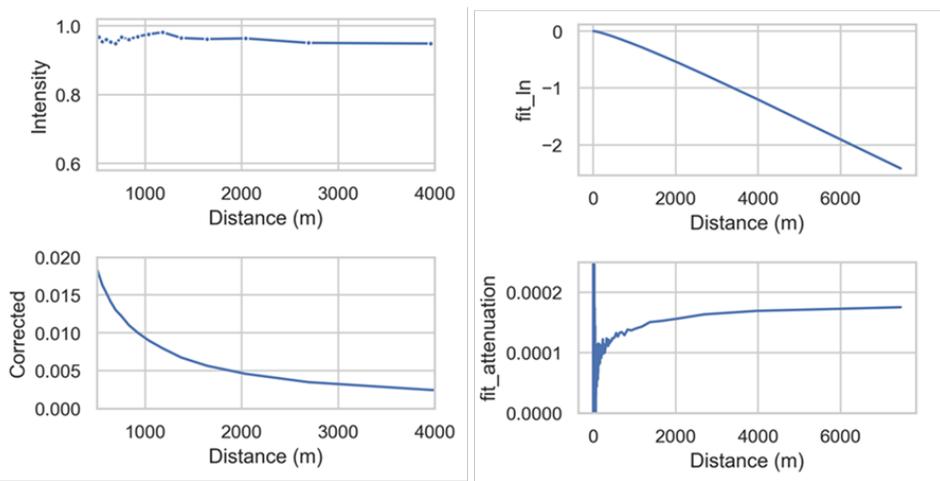

Figure 20: Raw and processed data from the UV lidar collection. (Top Left) Scaled raw intensity from each range column showing very flat dynamic range. (Bottom Left) Range corrected intensity using preliminary model. (Top Right) Natural log of the range corrected intensity. (Bottom Right) One-way extinction coefficient after applying the slope method.

Both systems reported a value of $0.00015\ m^{-1}$ in the furthest range, effectively at the end of the S-lidar's maximum range. The 355 nm S-lidar is able to make these measurements down to 10 m while the traditional ToF lidar is limited to 500 m. However, it was evident that an additional correction or more likely, a comprehensive correction and calibration method to be implemented. We believe that this correction appears as a linear range-dependent function applied to the $\ln(P)$ calculation.

## 4. CONCLUSIONS

In this work, we present novel theory and preliminary results to motivate the potential future applications of Scheimpflug technology as a tool for direct observation and ranging for several active and passive remote sensing applications. We highlight the unique potential of these systems to act as flexible and low SWaP-C alternatives to ToF lidars with particular benefits to near-range measurements of laser propagation. Through this work we were able to show Scheimpflug lidars have the following system-level benefits:

- There is intrinsically a lower dynamic range requirement since the typical range squared decrease in a lidar detector's power is offset by a range squared increase to the projected pixel size at the ideal object plane.
- Range resolution increases at nearer distances and ranging performance can be tuned with opto-mechanical components and zoom lenses.
- These systems can work with continuous wave (CW) lasers or pulsed lasers, and use any 1D or 2D sensor arrays.
- The technology will work at any wavelength provided there is a camera solution and a sufficient natural light source or available laser to create sufficient backscatter within an acceptable exposure time.

With our simultaneous experiment with the ToF lidar and imaging the same laser beam, we directly saw the value and performance capability of the 355 nm prototype Scheimpflug lidar. Having designed and built both systems, the S-lidar is much less complex both for the receiver design and overall system hardware. Without any models beyond the pixel-range map, it is able to create range profiles with enough sensitivity to visually observe the atmospheric effects from extinction and turbulence. Similarly, the smaller beam of the 532 nm prototype was extremely sensitive to turbulence and easily observable beam wander and scintillation effects occurring on the live video.

The significant remaining challenge is still developing a comprehensive inversion method or volume scattering model that accounts for depth of field and defocus. As a comparison to traditional lidar, this problem is relatable to the issue of having geometric overlap values less than unity.

## ACKNOWLEDGEMENTS

We would like to thank GTRI's Independent Research and Development (IRAD) funds for supporting this work. We are grateful to Amanda Villegas, Claudia Vitale, and Yassine Fouchal for supporting the earliest efforts of this research at Georgia Tech Research Institute. We also thank the team of Rossy Dang, Dillan Synan, Mikhail Velez, and Varun Vudathu for their inspiring research and generating our first Scheimpflug lidar point cloud. On behalf of all authors, the corresponding author states that there is no conflict of interest.